\documentstyle[prd,aps,preprint]{revtex}
\begin{document}
\draft
%
%  Uncomment following two lines and one below for 2 column format.
%@twocolumnfalse\endcsname

\preprint{Nisho-97/6} \title{Spontaneous Magnetization of Axion Domain Wall 
and Primordial Magnetic Field} 
\author{Aiichi Iwazaki}
\address{Department of Physics, Nishogakusha University, Shonan Ohi Chiba
  277,\ Japan.} \date{May 15, 1997} \maketitle
\begin{abstract}
We show that axion domain walls gain spontaneous
magnetization in early universe by 
trapping either electrons or positrons 
with their spins polarized.
The reason is that the walls produces an attractive potential for these 
particles.
We argue that the wall bounded by an axionic superconducting string
leaves a magnetic field after its decay.
We obtain a field $\sim 10^{-23}$ Gauss on the scale of horizon 
at the recombination.
\end{abstract}

\pacs{14.80.Mz, 98.80.Cq, 98.62.En 
% \\Axion, Domain Wall, Primordial 
%Magnetic Field 
\hspace*{3cm}}
\vskip2pc
%%%%%%%%%%%%%%%%%%%%%%%%%%%%%%%%%%%%%%%%%%%%%%%%

%%%%%%%%%%%%%%%%%%%%%%
Magnetic fields in galaxies or stars are present in our universe.
Nevertheless, their origin is still unknown although there are several
cosmological origins proposed\cite{qcd,ew,iw,in,etc}
; it could arise during quark-hadron phase 
transition\cite{qcd}, electro-weak phase transition\cite{ew,iw}, 
inflation\cite{in} e.t.c..
It is generally assumed that dynamo process\cite{dy} amplifies the seed of 
small primordial magnetic field $\sim10^{-18}$ Gauss generated by these origins
to the observed one $10^{-6}$ Gauss in galaxy.

Previously we have discussed\cite{iw} a magnetic field associated with 
domain walls, which gain ferromagnetism owing to  
fermion zero modes\cite{zerod} bounded to the walls. 
But the zero modes does not necessarily exist in any realistic models.
In this letter we show that axion domain walls, although they do not possess 
such zero modes, become ferromagnetic 
by trapping either electrons or positrons 
in the hot universe. The walls leave dipole magnetic fields after their decay
with various strengths and sizes, 
e.g. a field of $\sim 10^{20}$ Gauss 
on scales of $\sim10^{-3}$ cm at the temperature $\sim100$ MeV. 
These dipole fields generate a large scale magnetic field. 
The field on the scale of horizon is estimated to possess 
its strength $10^{-23}$ Gauss at the recombination of photons and electrons.
This magnetic field is a candidate of primordial magnetic fields which 
lead to galactic or intergalactic
magnetic field at present.

Let us begin to briefly explain axion domain walls\cite{kim}.
The axion is a Goldstone mode\cite{axion} of 
Pecci-Quinn global U(1) symmetry\cite{PQ}
which is broken spontaneously with the energy scale 
$f_{PQ}$, $10^{10}\sim 10^{12}$ GeV. 
This massless axion gains a mass $m_a$ through effects of QCD instantons 
due to anomaly\cite{t'Hooft} of the symmetry. 
Since the axion field $a$ is essentially the 
phase $\theta$ of a Higgs field $\sigma\sim f_{PQ}e^{i\theta}$ 
( $a=f_{PQ}\theta$ ), the vacuum $\theta=0$ of the axion is degenerate with
the vacuum $\theta=2\pi$. This fact leads to 
domain walls between these degenerate vacua. Namely the domain walls 
are produced during QCD phase transition 
at which QCD instantons work effectively.
Besides the domain wall solitons, string solitons associated with
the breaking of the $U(1)$ symmetry are produced at the temperature $f_{PQ}$. 
In general they are superconducting\cite{super}.
Since we consider an axion model 
with the color anomaly of the Pecci-Quinn symmetry equal to one,
domain walls are surrounded by these strings.
Furthermore, they 
decay soon after their production without dominating the universe.
As we show below, however, the walls leave magnetic fields with various sizes
and strengths. This is because the walls gain magnetic moments by trapping 
either electrons or positrons with their spins polarized.
It turns out that in the decay of the walls 
the superconducting strings play important roles 
on producing the magnetic fields.
( For definiteness we use numerical values such that 
$f_{PQ}=10^{12}$ GeV and $m_a=10^{-5}$ eV 
throughout the paper. )

In order to see the spontaneous magnetization of the wall
we first show that the axion domain wall generates an attractive potential
for electrons and positrons with a particular polarization.  
Suppose an axial vector coupling between electron field $\psi$ 
and axion field $a$, 

\begin{equation}
L_{int}=g\partial_{\mu}a \bar{\psi}\gamma_5\gamma^{\mu}\psi
\label{g}
\end{equation}
where $g$ is the coupling constant whose value is given by $c f_{PQ}^{-1}$,
with a numerical positive constant $c$; the value of $c$ depends on models
of the axion.
Then, we can see easily that the wall produce a spin dependent potential for
electrons and positrons. Assuming that the flat wall is located at 
$x_3=0$, 
we rewrite Dirac equation of $\psi= {u\choose v}$ such that 

\begin{equation}
\varepsilon u = (-\frac{\partial^2}{2m} + g\partial_3a\sigma_3)u
\quad 
\left(\ \varepsilon v = (-\frac{\partial^2}{2m} - 
g\partial_3a\sigma_3)v\ \right)
\end{equation}
with energy $E=m+\varepsilon$ ( $E=-m-\varepsilon$ ); 
$m$ is the mass of electron.
We have taken the nonrelativistic limit so that 
only large component $u$ ( $v$ ) is exhibited; small component $v$ ( $u$ )
is given by $v=-i\partial_3u/2m$ ( $u=i\partial_3v/2m$ ).

We note that the typical scale of spatial variation of $\theta$ is given 
by the axion mass $m_a$; it increases from $0$ to $2\pi$ along  
$x_3$ axis.
Thus it turns out that 
the potential $g\partial_3a\sigma_3$ 
for both of electrons and positrons
is attractive for states with spin down,
and repulsive for states with spin up. It has approximately 
the width of $m_a^{-1}\sim 1$ cm and the depth of $m_a\sim 10^{-5}$ eV
when the constant of $c$ is order of one.
Hence the potential depth is quite shallow but
the width is so large that the potential can 
accommodate bound states for electrons and positrons. 
Obviously, spins of these particles bounded to the wall are aligned.
This causes the ferromagnetism of the axion domain wall.

It seems that in the universe with its temperature $\sim 100$ MeV 
the states with such small binding energies are irrelevant 
for the property of the wall. But 
we should note that there exists a small fraction of the particles 
occupying the bound states
even in such hot universe; there is 
tiny but nonvanishing probability of
a particle occupying the state. This small fraction of the particles 
gives rise to the magnetic property of the wall 
whose relevant energy scale is
much smaller than the temperature. 

Secondly we show that for a strong magnetic field either electrons or positrons
have no bound states even in such attractive potential. For the purpose  
let us see the energy spectra of these particles with 
spins down under the magnetic field imposed perpendicular to the wall.
When the field is pointed to the direction 
along positive $x_3$ axis, the binding energies of these particles are 
found easily,

\begin{eqnarray}
\varepsilon_{k,n}&=&-m_a + \frac{k_3^2}{2m} + 
\omega n\quad  \mbox{for electron}\nonumber\\
\varepsilon_{k,n+1}&=&-m_a + \frac{k_3^2}{2m} +\omega (n+1)\quad 
 \mbox{for positron}
\end{eqnarray}
where $n$ is integer ( $\geq 0$ ) and 
$\omega$ is the cyclotron frequency
( $\omega=eB/m$ ); $k_3$ is momentum along $x_3$ axis. 
For convenience we have simplified the attractive potential by a square well 
potential with a width equal to $m_a^{-1}$ and a depth equal to $m_a$.
The difference in the energy spectra 
between electrons and positrons comes from the difference in 
the directions of the magnetic moments of these particles in the 
attractive potential of the wall.

We note that when binding energies are positive ( $\varepsilon \geq 0$ ),
the particles are not bounded. Such localized states are quite unstable 
against any small perturbations, e.g. couplings with scattering states of the 
particles, oscillation modes of the wall e.t.c..
Especially thermal fluctuations would destroy such unstable states.
Thus the states with $\varepsilon >0$ must decay.
Hence for the sufficiently strong magnetic field
$eB\geq m_am$, positrons do not form bound states on the wall since
the lowest energy $\varepsilon_{k=0,1}$ ( $=-m_a+\omega$ ) 
of positrons becomes positive.
Similarly, when the magnetic field is pointed to the direction along 
the negative $x_3$ axis, electrons do not form bound states.

If the magnetic field is generated by these polarized 
electrons or positrons themselves bounded to the wall, it means that 
the wall gains magnetization 
spontaneously. As will be shown, when the temperature is higher than 
$180$ MeV, the wall gains the spontaneous magnetization 
with sufficiently strong magnetic field.

Now we show the spontaneous magnetization of the wall 
by calculating free energy of the gas of 
electrons bounded to the wall. The free energy $F_M$ of the system 
under the magnetic field is defined
as 

\begin{equation}
F_M=-\beta^{-1}\sum_{\varepsilon_{k,n}\leq 0 }
\log{(1+e^{-(m+\varepsilon_{k,n})\beta})}
\end{equation}
where we have taken account of only contribution 
of electrons bounded to the wall, since positrons 
have no bound states under the sufficiently strong 
magnetic field. 
Summation, $\sum_{\varepsilon_{k,n}\leq 0 }$  
implies $N_d\sum_{n=0}\int m_a^{-1}dk_3/(2\pi)$ 
with the condition 
that $\varepsilon_{k,n}\leq 0$; $N_d$ is the degeneracy of each Landau level
given by $eBS/2\pi$ where $S$ is the surface area of the wall. 
$\beta^{-1}$ is the temperature of the universe.
We have assumed the chemical potential to be vanishingly small.
Noting that only states relevant in the summation are states of the lowest 
Landau level, we find that 
 $F_M$ is approximately given by 

\begin{equation}
F_M=-\frac{m_a^{-1}N_d2\log2\\\ \sqrt{2mm_a}}{2\pi\beta}
 =-\frac{eBS\log2\\\ \sqrt{2m/m_a}}{2\pi^2\beta}
\label{f}
\end{equation}
in the limit of the temperature being much higher than the mass of electron,
$\beta^{-1}>> m$.
Thus magnetization of the wall is 

\begin{equation}
M=-\frac{m_a}{S}\frac{\partial F_M}{\partial B}
=\frac{e\log2\\\ \sqrt{2mm_a}}{2\pi^2\beta}
\end{equation}
This magnetization induces a current $J$ at the boundary of the wall;
$J=M/m_a$. Then, the current induces a magnetic field which is not 
uniform; but it is approximated to  
$B_0=J/R$ where $R$ is the radius of the wall. 
We identify this magnetic field 
$B_0$ with the field $B$ in the free energy $F_M$.
In this way we obtain the magnetic field generated 
by polarized electrons bounded to the wall,  

\begin{equation}
B=\frac{e\sqrt{2}\log2}{2\pi^2R\beta}\sqrt{\frac{m}{m_a}}
\label{b}
\end{equation}

For the consistency this magnetic field has to satisfy a condition, 
$eB\geq mm_a$.
Otherwise the assumption that only 
particles bounded to the wall are electrons and 
that positrons have no contribution to the free energy, does not hold.
Therefore when the condition is satisfied, the wall may gain the spontaneous
magnetization by trapping either electrons or positrons.

%It seems apparently that all of electrons can not be bounded to the wall
%under such high temperature $\sim 100$ MeV 
%and the bound states are irrelevant. 
%We should mention, however, that although the temperature is much higher than 
%the binding energies of electrons, small fraction of the particles 
%occupies the bound states and determines the magnetic property of the wall. 
%The magnetization $M$ is the one generated 
%by the small fraction of these particles.

As have been shown in numerical simulations\cite{dm}, 
energetically important domain walls 
are ones with boundaries of horizon size. Thus we equate $R$ 
with the distance to the horizon. Noting that $R$ is about equal to 
$0.3M_{pl}\beta^2/\sqrt{f}$ ( $M_{pl}$ is Planck mass and $f$ is 
dynamical massless degrees of freedom at temperature $\beta^{-1}$ )\cite{u}, 
we find from eq(\ref{b}) that 
the consistency condition,

\begin{equation}
B\geq mm_a/e\sim 230\mbox{Gauss}
\label{cond}
\end{equation} 
is satisfied when the temperature is higher than $180$ MeV. 
Therefore the wall gain the magnetization $M$ spontaneously 
when it is produced during QCD phase transition ( $\sim 200$ MeV ).
Note that $m_a$ in the above formulae represents the depth of the potential
when the coupling constant $g$ in eq(\ref{g}) between the axion and electron
is the order of $f_{PQ}^{-1}$.
On the other hand if $g$ is the order of $\alpha g$ 
( $\alpha=e^2/4\pi$ ) as predicted 
in a model of a hadronic axion\cite{kim}, 
then the depth of the potential $m_a$ is replaced by
$\alpha m_a$. Thus the minimum temperature needed 
for the appearance of the magnetization is given approximately by $15$ MeV.

In order to establish the fact that the spontaneous magnetization actually 
arises, we have to make sure that the above free energy $F_M$ is 
lower than that of the state with no magnetization. 
Such a free energy is given by

\begin{equation}
F_0=-2\beta^{-1}\sum_{\varepsilon_k\leq 0 }\log(1+e^{-(m+\varepsilon_k)\beta})
\sim -2 \beta^{-1}\log2\sum_{\varepsilon_k\leq 0}
=-\frac{2S\log{2}\\\ mm_a\sqrt{2mm_a}}{3m_a\pi^2\beta}
\end{equation}
with $\varepsilon_k=-m_a+\vec{k}^2/2m$, 
where we have taken both contributions 
from electrons and positrons, and have taken the limit of the temperature 
being much larger than $m$; $\vec{k}$ is 3-momentum.
Comparing this free energy $F_0$ with the above one $F_M$ 
in eq(\ref{f}) we find that $F_0>F_M$
when a condition, $eB>4mm_a/3$, is satisfied; the condition is
roughly the same as the above condition eq(\ref{cond}).
Therefore spontaneous magnetization of the axion domain wall 
of horizon size arises when the temperature is higher 
than about $180$ MeV. In the case the magnetic field stronger 
than $200$ Gauss is generated.

It seems apparently strange that the spontaneous magnetization arises 
in the side of higher temperature than the critical one. In general, ordered 
states appear in the side of lower temperature than the critical one,
possessing smaller entropies than those of disordered states.   
In our case, however, the entropy of the state with the magnetization 
is larger than that of the state without the magnetization, 
when $B$ is larger than a critical value given in eq(\ref{cond}).
This is because the number of the bound states, especially the states of the 
lowest Landau level
increase with the magnetic field ( the degeneracy $N_d$ increases with $B$ )
and consequently the entropy becomes large.
This magnetic field is proportional to the magnetization which increases with
the temperature: The thermal fluctuation leads to large orbital 
angular momentum and hence large magnetic moment. 
( Note that states with various angular momenta are degenerate 
in a Landau level. )
Thus when the temperature 
becomes large, the magnetic field increases and the entropy becomes large. 
Eventually it dominates over the entropy of the state 
without the magnetization and 
consequently the spontaneous magnetization can arise.

As we have shown, the axion domain walls trap either electrons or positrons
with their spins polarized and gain the magnetization. Although  
they also gain electric charges, the charges are screened immediately
due to large electric conductivity of the universe. Thus charge neutrality
is kept.

Finally we discuss magnetic fields left after the decay of the walls.
We easily understand that during the decay of the walls, the magnetic field 
becomes strong as the radius of the walls decreases.
To see it we note that the walls are surrounded by superconducting strings
which must carries the boundary current $J=M/m_a$. This current is supposed 
to be carried by fermion zero modes\cite{zeros,super} on the strings. 
Then as the number of the zero modes 
( $\sim JR$ ) is conserved, the current increases as the radius $R$ of 
the string decreases.
The strings shrink with increasing the current until 
the current is saturated\cite{super} with its maximal value 
$qm_f/\pi$; $q$ and $m_f$ are the electric charge and the mass 
of the fermion making zero modes on the strings.
Subsequently, the strings shrink without increasing the current 
but by emitting the fermion zero modes. 
Therefore, when we start with, for example, a magnetic field $10^2$ Gauss 
of the walls on the scale of horizon $R_h\sim 10^6$ cm 
at the temperature $100$ MeV, 
the field $B_c=10^{2}(R_h/L_c)^2\sim 10^{20}$ Gauss 
on the scale of $L_c$ is achieved when the current is saturated with the 
radius of the strings being equal to $L_c$. 
$L_c$ is determined by $MR_h/m_a=qm_fL_c/\pi$ 
( conservation of the zero modes ); $L_c\sim 10^{-3}$ cm with 
use of $m_f=10^{12}$ GeV and $q=1$ 
( this is a typical energy scale $f_{PQ}$ of 
the fermion mass which is generated through Higgs mechanism 
associated with Pecci-Quinn symmetry ). 
For smaller radius R of the strings, 
stronger magnetic fields ( $\propto 1/R$ ) arise on smaller scales of $R$.
On the other hand the large scale magnetic field should be determined by 
these randomly oriented dipole fields. Thus they produce a 
magnetic field $B(L)$ on the scale of $L$ such that
$B(L)=B_c(L_c/L)^{3/2}$\cite{ls,qcd,ew}. The field evolves to 
the field $B_{re}(L_{re})=B(L)(1\mbox{eV}/100\mbox{MeV})^2$ 
on the scale of $L_{re}=L(100\mbox{MeV}/1\mbox{eV})$
at the recombination ( $1$ eV ). Note that the magnetic flux is conserved 
in the early universe owing to large electric conductivity; 
$Ba_{RW}^2$ = constant ( $a_{RW}$ is 
the cosmic scale factor in Robertson-Walker metric ). 
As we are concerned with the field on scales 
of horizon size $L_h\sim 0.1$Mpc at the recombination,
we find that $B_{re}(L_h)=10^{-23}$ Gauss. 
Similarly we obtain the magnetic field $\sim10^{8}$ Gauss
on scales of $\sim 10^4$ cm at the nucleosynthesis. This is sufficiently
small not to affect seriously the production of light elements\cite{nuc}.   
In this way the ferromagnetic axion domain walls generates the primordial 
magnetic fields.

\begin{flushleft}
Acknowledgments

The author would like to thank Prof. M. Kawasaki 
for useful discussions and staff members of theory division in
Institute for Nuclear Study, University of Tokyo for their
hospitality.
\end{flushleft}

%\begin{figure}
%\caption{
%  The solid ( dashed ) curve represents $I-V$ characteristic with
% $\beta = 0$ ( $\beta = 4$ ) in DC current feed with normalized
%  voltage $V/R_qI_c$ in the horizontal axis.  The effect of the
%  capacitance ($\beta \ne 0$ ) leads to a hysteresis; the voltage
%  raises linealy with current $I$ ($\leq I_c$ ), but suddenly jumps (
%  indicated with the dotted line ) when the current reaches at the
%  critical current $I_c$. Then it follows the dashed curve with the
%  current increasing furthermore. Conversely, the voltage decreases
%  continuously following the dashed curve without passing the dotted
%  line even if the current becomes less than $I_c$.  }
%\end{figure}

%\begin{figure}
%\caption{
%  The dotted curve shows Shapiro-like steps ( voltage jumps in the
%  unit of $\omega /e$ ) in $I-V$ characteristic with AC current feed;
%  its frequency $\omega$ is assumed to be $0.16 e R_q I_c$.  }
%\end{figure}

%\begin{figure}
%\caption{The solid curve represents $I-V$ characterisic in 
%  DC voltage feed with $R_q/R =2$. The dotted curve shows Shapiro-like
%  steps ( current drops in the unit of $\omega/eR$ ) in $I-V$
%  characteristic with AC voltage feed; its frequency $\omeis
%  assumed to be $0.16 e V_c R_q /( R + R_q )$.  }
%\end{figure}

\end{document}